%% file: kril.tex
\documentclass[12pt]{iopart}
\usepackage{iopams}
\usepackage{epsfig}

\def\ead#1{\vspace*{5pt}\address{E-mail: \mailto{#1}}}
\def\mailto#1{{\tt #1}}


 \newcommand{\bc}{\begin{center}}
\def\ec{\end{center}}

\begin{document}
\title{Experimental approaches to low $x$ at HERA}  
\author{J\"org Gayler}

\address{DESY, Notkestrasse 85, 22603 Hamburg, Germany}
\ead{gayler@mail.desy.de}

\begin{abstract} 
 Data are presented on the production of jets and $\pi^0$ mesons
 at low Bjorken $x$ in a kinematic region
 where
 standard DGLAP evolution in $Q^2$ gives little   
 phase space for high $p_t$ particle and jet production.
 The data are compared with various QCD models 
 based on different treatments of parton emissions at small $x$.
\end{abstract} 

\section{Introduction}
\begin{figure}[ht]
\begin{center}
\begin{picture}(200,110)
\put(-30.,1.)
{\epsfig{file=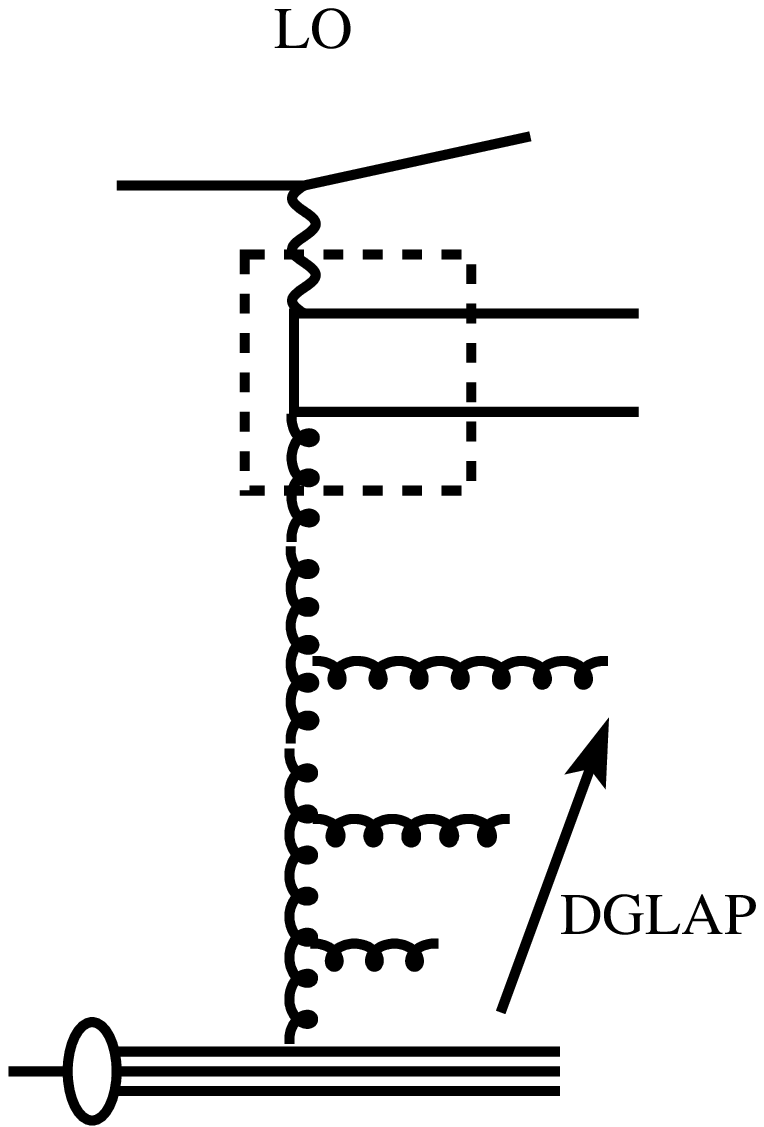,width=75pt}}
\put(70.,0.)
{\epsfig{file=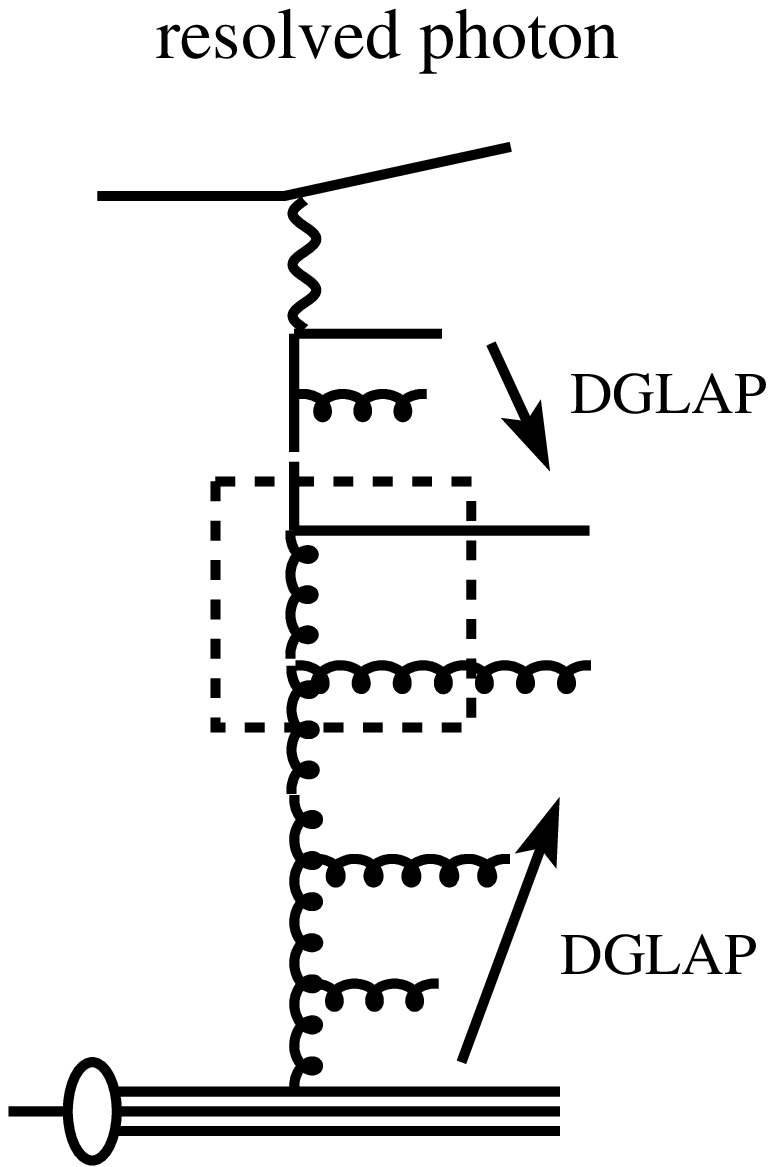,width=75pt}}
\put(167.,0.)
{\epsfig{file=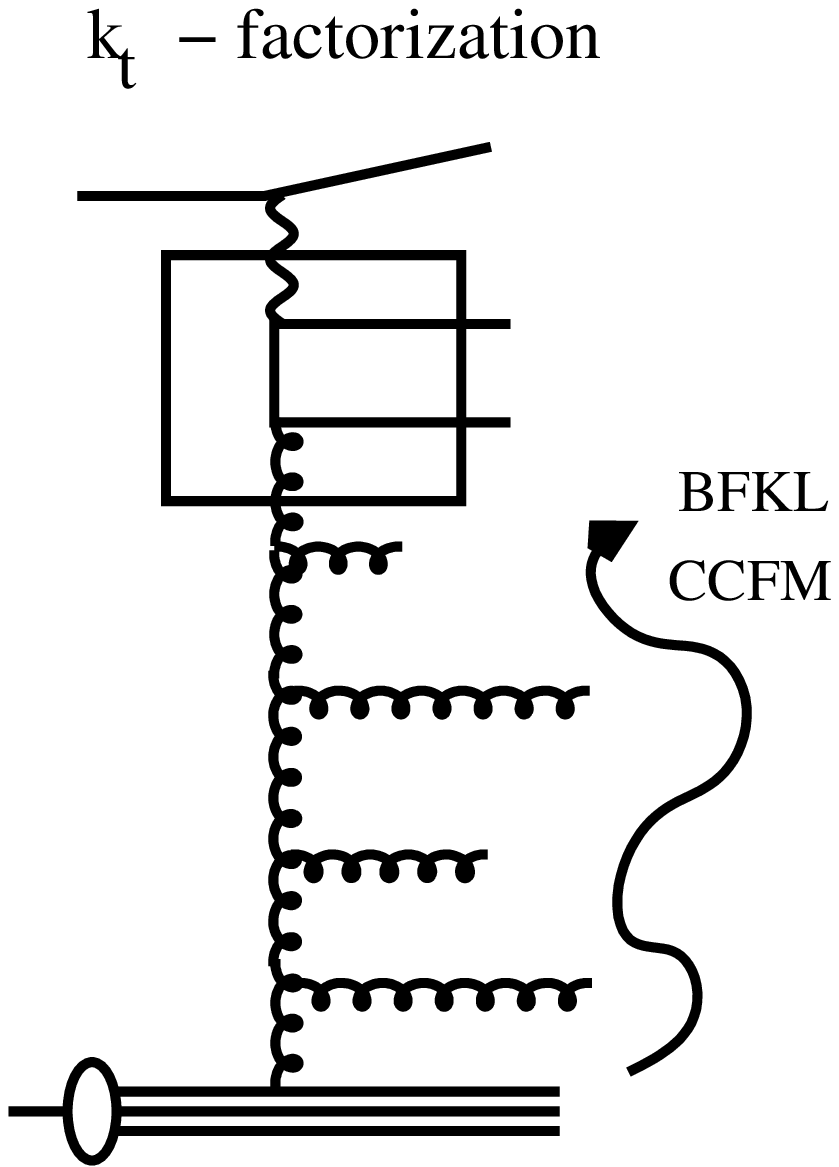,width=80pt}}
 \end{picture}
\caption{
Different evolution schemes. Left: Direct photon coupling to leading
order matrix element of the hard process. The arrow indicates the ordering in
 increasing $k_t$
in DGLAP evolution. Center: same, but for resolved photons.
Right: BFKL and CCFM approach with ordering in energy and angle
    respectively. 
\label{fig:ladder}}
\end{center}
\end{figure}

 Experimentally
    QCD dynamics at low $x$ are mostly studied in nucleon structure
    functions (for recent results see~\cite{Kappes:2002dz,jg}),
   heavy quark production in deep inelastic scattering (DIS)~\cite{zotov},
   and forward jets and particle production (see also~\cite{merino}).
 Results on the latter
   are presented here\footnote{Presented at XXXII
   International Symposium on Multiparticle Dynamics,
   Alushta, Crimea,  \\
 7. - 13th September 2002}.
 ``Forward'' refers in the present context
   of {\it ep} interactions to the
   region close to the outgoing proton beam.
   This region is particularly interesting, as
 the large energies
 available at small $x$ give rise to
  a large phase space for gluon ladders which are sensitive to different
 QCD evolution schemes 
 as indicated in 
  Fig.~\ref{fig:ladder}.
Due to the increasing virtualities $k_t$ 
towards the hard interaction in the DGLAP~\cite{DGLAP}
 evolution scheme
(Fig.~\ref{fig:ladder},~left),
large energy jets and hadrons with substantial  $p_t$ are suppressed in
 the forward direction.
 If a resolved hadronic photon structure is considered,
 this strict ordering is broken
(Fig.~\ref{fig:ladder},~centre).
 Finally, based on the BFKL~\cite{bfkl} (CCFM)~\cite{ccfm} equation,
ordering in energy (angle) is obtained with $k_t$ factorisation.
 In this case substantial $p_t$ can be 
 expected anywhere in the ladder
(Fig.~\ref{fig:ladder},~right).   
 
 NLO pQCD calculations to order $\alpha_s^2$ are not able 
 to describe forward jet data at small $Q^2$ where the NLO corrections
 are very large~\cite{Adloff:2002ew}.

 In this report data~\cite{H1ichep}
 on forward jet and $\pi^0$ production
 (see ref.~\cite{prev} for previous data)
 are compared with several QCD based Monte Carlo (MC)
 models.
 The RAPGAP MC model~\cite{Jung:1993gf}
 combines leading order (LO) matrix
 elements with DGLAP parton showers and is used with and without
 resolved virtual photon contributions.
 The treatment of higher orders in the ARIADNE program~\cite{Lonnblad:1992tz},
 which is based on the Colour Dipole Model (CDM)~\cite{Andersson:1989ki},
 leads to unordered parton emissions, as expected in the BFKL approach.
 Finally, CASCADE~\cite{Jung:1998mi} corresponds to a solution of the CCFM equation.

\section{Results}
\begin{figure}[ht]
\begin{picture}(200,240)
\put(0.,-12.)
{\epsfig{file=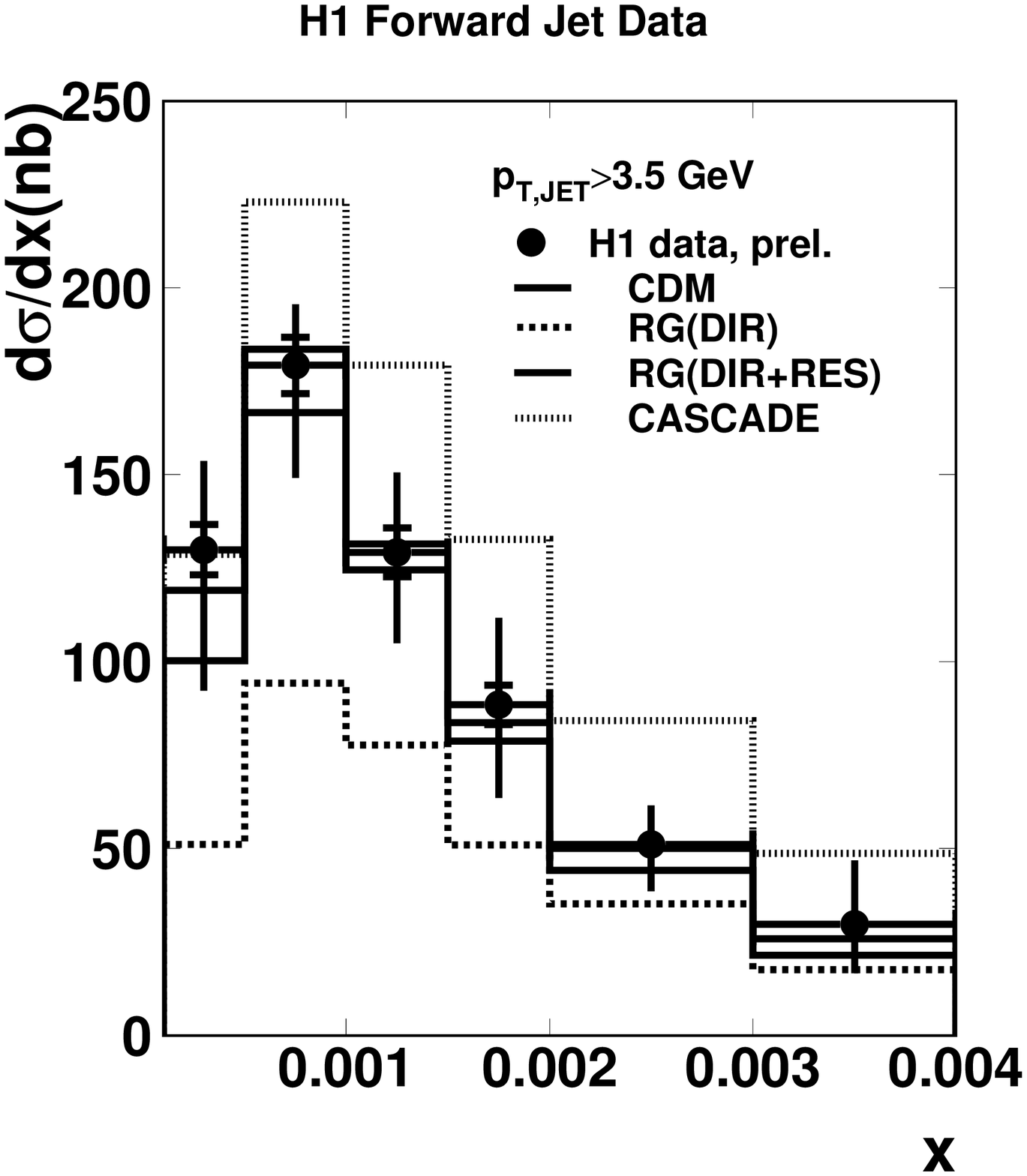,width=230pt}}
\put(224.,-2.)
{\epsfig{file=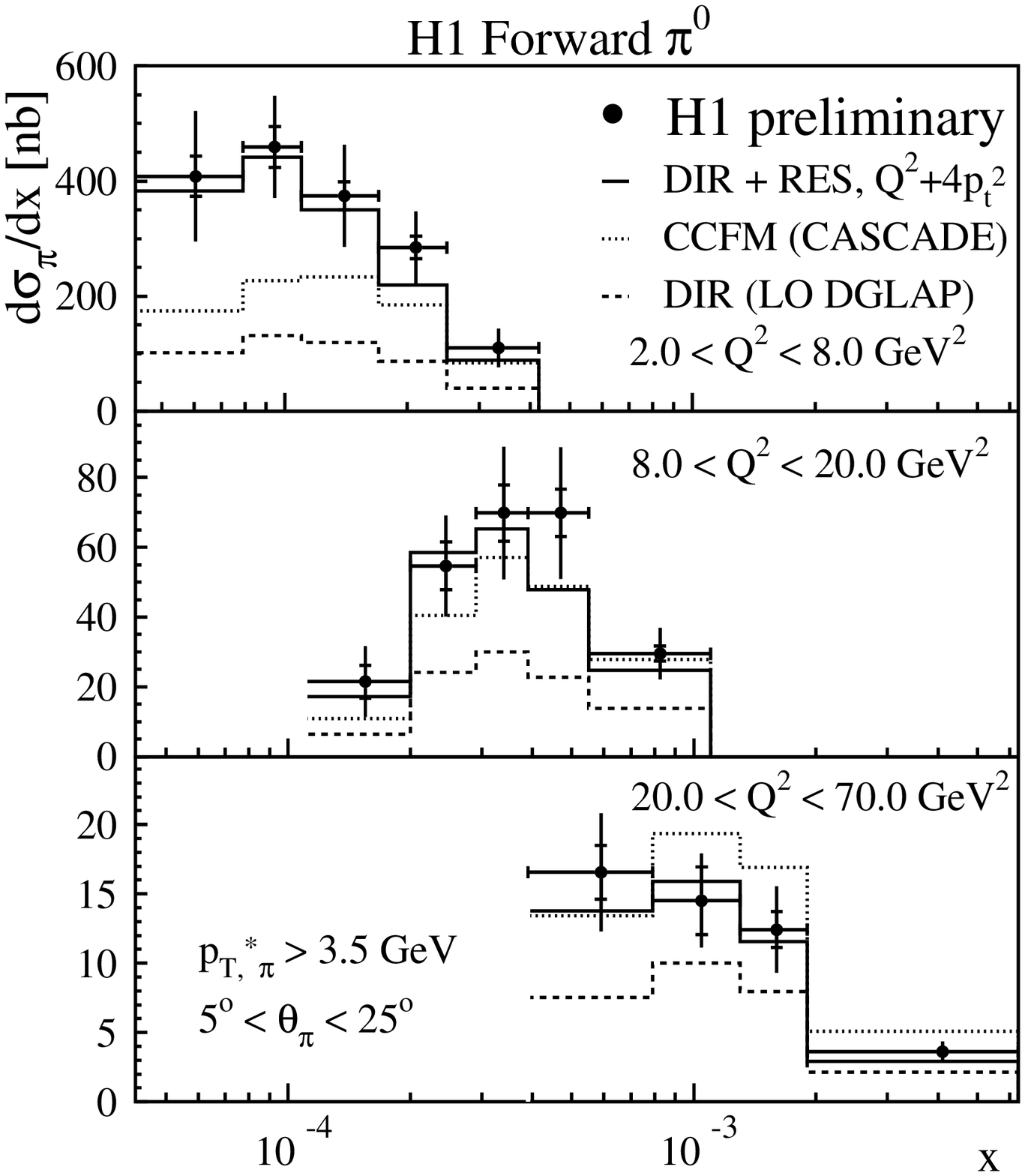,width=225pt}}
 \end{picture}
\caption{
  Forward jet (left) and forward $\pi^0$ (right) data with
  predictions of RAPGAP with direct and resolved $\gamma$ interactions 
  and DGLAP LO parton showers, 
  ARIADNE based on the Colour Dipole Model (CDM) and CASCADE based
  on the CCFM equation.
\label{fig:data}}
\end{figure}

Jets are selected using the inclusive $k_t$ algorithm in the kinematic range
 $5 < Q^2 < 75$ GeV$^2$,
 $7 <  \theta_{jet} < 20^{\circ}$.
 Substantial jet $p_t$ and jet energy
  is required by the conditions  $0.5 < p_{t\; jet}^2/Q^2 < 2$      
  and  $x_{jet} = E_{jet}/E_p > 0.035$, where $E_p$ is the incident
  proton energy. The $\pi^0$ mesons are selected in a similar kinematic
  range with transverse momentum in the hadronic centre of mass system (CMS)
  $p^*_{t,\pi} > 2.5$ or $> 3.5$ GeV. The results in Fig.~\ref{fig:data}
show that in both cases the DGLAP RAPGAP model with only direct photon 
 interactions is well below the data. However, a very good description
 is achieved if resolved photon interactions are included.
 Also the CDM model describes the jet data very well.
 The CCFM CASCADE model predicts
 cross sections that are too large at large $x$ $(x \gtrsim 0.001)$.

\begin{figure}[ht]
\begin{center}
\begin{picture}(200,417)
\put(-80.,0.)
{\epsfig{file=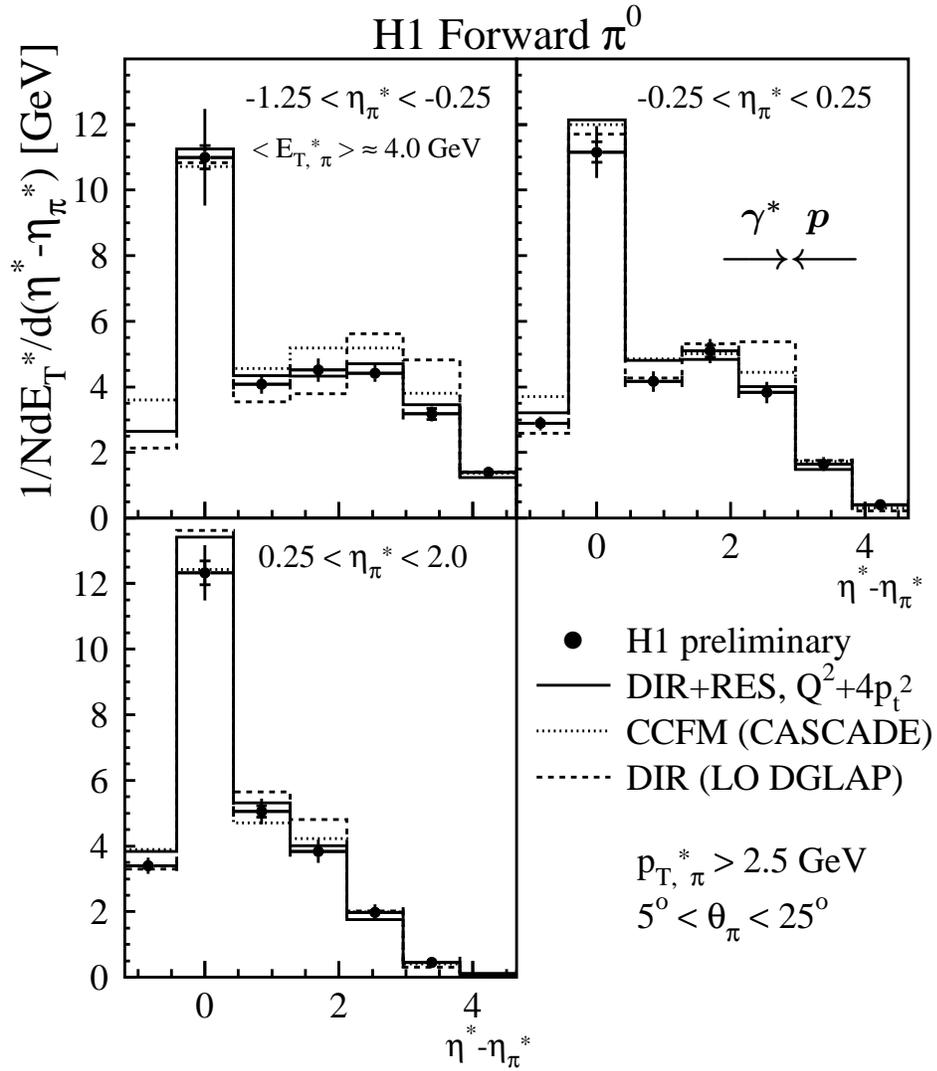,width=409pt}}
 \end{picture}
\caption{
Transverse energy flow relative to the $\pi^0$
 for different ranges of pseudo rapidity $\eta_{\pi^*}$
in the hadronic CMS compared with the predictions of CASCADE,
and RAPGAP with DGLAP parton showers with (DIR+RES) and without (DIR)
 resolved virtual photon interactions. The contribution from the $\pi^0$
 itself is included in the energy flow.    
\label{fig:flow}}
\end{center}
\end{figure}

 It is interesting to see (Fig.~\ref{fig:flow}), that 
 the transverse energy flow between the photon coupling and the $\pi^0$
 meson
 also does not follow the expectations from 
direct photon interactions with
 DGLAP parton showers, which predict a larger amount of
 $E_t$ close
to the virtual photon than is observed in the data.

{\large
\vspace*{102pt} \hspace*{330pt} {\boldmath       $\gamma^* \;\; p$}   

  \hspace*{322pt} {\boldmath $\longrightarrow \longleftarrow$} }
   
\vspace*{-124pt} 

\section{Conclusion}
 The presented DIS forward jet and $\pi^0$ data can be described
 by LO matrix elements and DGLAP parton showers only if resolved 
 photon interactions are included.
 The CDM model, which contains photon emissions unordered in $k_t$,
 describes the jet data very well. The CASCADE MC does not
 describe the $x$ dependence of the data.
         
\section*{Acknowledgments}
   I am grateful to Hannes Jung, Lidia Goerlich, Martin Karlsson,
  Paul Newman,
  Grazyna Nowak and Jacek Turnau for discussions or comments.
  I thank the organisers for an interesting conference, in particular
  Elena Kolganova for kind support.

\end{document}






%% file: kril.bbl
\section*{References}     
\begin{thebibliography}{99}

\bibitem{Kappes:2002dz}
A.~Kappes  [ZEUS Collaboration],
``Structure function results from ZEUS'',
hep-ex/0210032,     \\
proceedings ICHEP 2002, Amsterdam;  \\
\hspace*{-14pt} Z.~Zhang [H1 Collaboration]
``Structure function results from H1'', ibid.

\bibitem{jg}
J. Gayler,
``Proton Structure Functions Measurements from HERA'',
 these proceedings,   \\ 
  hep-ex/0211051,   

\bibitem{zotov}
N. Zotov,
  ``Heavy quark production with BFKL and CCFM Dynamics'', these proceedings;
\hspace*{-14pt} S.~P.~Baranov, H.~Jung, L.~Jonsson, S.~Padhi and N.~P.~Zotov,
Eur.\ Phys.\ J.\ C {\bf 24} (2002) 425.
 
\bibitem{merino}
C. Merino,
``Correlation between average pt and jet multiplicities from the BFKL
        pomeron'', these proceedings.

\bibitem{DGLAP}
Y.~L.~Dokshitzer,
Sov.\ Phys.\ JETP {\bf 46} (1977) 641;    \\
\hspace*{-14pt} V.~N.~Gribov     
    and L.~N.~Lipatov,
Yad.\ Fiz.\  {\bf 15} (1972) 1218;
Yad.\ Fiz.\  {\bf 15} (1972) 781; \\
\hspace*{-14pt} G.~Altarelli and G.~Parisi,
Nucl.\ Phys.\ B {\bf 126} (1977) 298.
 
\bibitem{bfkl}
E.~A.~Kuraev, L.~N.~Lipatov and V.~S.~Fadin,
Sov.\ Phys.\ JETP {\bf 44} (1976) 443;       \\
Sov.\ Phys.\ JETP {\bf 45} (1977) 199;  \\
\hspace*{-14pt} I.~I.~Balitsky and L.~N.~Lipatov,
Sov.\ J.\ Nucl.\ Phys.\  {\bf 28} (1978) 822.
 
\bibitem{ccfm}
M.~Ciafaloni,
Nucl.\ Phys.\ B {\bf 296} (1988) 49;   \\
\hspace*{-14pt} S.~Catani, F.~Fiorani and G.~Marchesini,
Phys.\ Lett.\ B {\bf 234} (1990) 339, 
Nucl.\ Phys.\ B {\bf 336}(1990)18;
\hspace*{-14pt} G.~Marchesini,
Nucl.\ Phys.\ B {\bf 445} (1995) 49.

\bibitem{Adloff:2002ew}
C.~Adloff {\it et al.}  [H1 Collaboration],
Phys.\ Lett.\ B {\bf 542} (2002) 193.

\bibitem{H1ichep}
H1 Collab., contributions to ICHEP 2002, Abstracts 1000
and 1001.

\bibitem{prev}
C.~Adloff {\it et al.}  [H1 Collab.],
Nucl.\ Phys.\ B {\bf 538} (1999) 3;
Phys.\ Lett.\ B {\bf 462} (1999) 440;   \\
 \hspace*{-14pt} J.~Breitweg {\it et al.}  [ZEUS Collab.],
Phys.\ Lett.\ B {\bf 474} (2000) 223.

\bibitem{Jung:1993gf}
H.~Jung,
Comput.\ Phys.\ Commun.\  {\bf 86} (1995) 147.


\bibitem{Lonnblad:1992tz}
L.~Lonnblad,
Comput.\ Phys.\ Commun.\  {\bf 71} (1992) 15.

\bibitem{Andersson:1989ki}
B.~Andersson, G.~Gustafson and L.~Lonnblad,
Nucl.\ Phys.\ B {\bf 339} (1990) 393.

\bibitem{Jung:1998mi}
H.~Jung,
arXiv:hep-ph/9908497.
\end{thebibliography}
